# SENONE-AWARE ADVERSARIAL MULTI-TASK TRAINING FOR UNSUPERVISED CHILD TO ADULT SPEECH ADAPTATION

*Richeng Duan, Nancy F. Chen*

Institute for Infocomm Research, A*STAR, Singapore

## ABSTRACT

Acoustic modeling for child speech is challenging due to the high acoustic variability caused by physiological differences in the vocal tract. The dearth of publicly available datasets makes the task more challenging. In this work, we propose a feature adaptation approach by exploiting adversarial multi-task training to minimize acoustic mismatch at the senone (tied triphone states) level between adult and child speech and leverage large amounts of transcribed adult speech. We validate the proposed method on three tasks: child speech recognition, child pronunciation assessment and child fluency score prediction. Empirical results indicate that our proposed approach consistently outperforms competitive baselines, achieving 7.7% relative error reduction on speech recognition and up to 25.2% relative gains on the evaluation tasks.

## 1. INTRODUCTION

Despite research efforts for more than two decades, it is still challenging to develop speech technology for children (e.g. recognize child speech in voice-activated applications and automatic child speech evaluation) [1, 2, 3, 4, 5, 6, 7, 8]. Acoustic modeling, a key component in speech technology, is challenging when applied to child speech. The challenges stem from physiological differences of the articulatory apparatus size, pronunciation variations, and proficiency levels between children and adults [9, 10, 11, 12]. While large-scale linguistic resources and powerful computational models enable the development of superior acoustic models, such privileged scenarios are often unavailable when processing child speech [13]. To overcome challenges of acoustic modeling for child speech and the scarcity of annotated linguistic resources, our previous work [14] proposed to learn a feature adaptation model to transform child speech to the adult feature space without using transcribed child speech. Similar to recent work of learning domain invariant feature representations [15, 16, 17, 18], [14] conducted adversarial training using binary domain labels (of child and adult). However, such training strategies implicitly assume the acoustic transformations needed for each phoneme is the same, while in practice the acoustic differences across adults and children vary across phonemes; resonant frequencies of vocal tract models of voiceless fricatives are more similar across adults and children while those of high vowels would be more distinct [19], making it more difficult for global transformations to achieve optimal results. Inspired by such linguistic insights, we propose a senone-aware adversarial training (SAT) strategy to adapt child speech to adult speech. The traditional binary child/adult domain labels are further elaborated and refined with multi-dimensional senone posterior labels to take advantage of phonemic differences across adults and children. Empirical validation on 3 tasks compares favorably with established baselines.

## 2. RELATED WORK

### 2.1. Acoustic modeling of children's speech

Training an acoustic model directly with thousands of hours of child speech is the most straightforward way to reach good performance [2]. However, this scenario is often unavailable. To improve child speech recognition and automatic spoken language assessment, past approaches include statistical machine learning such as vocal tract length normalization (VTLN) [20] and maximum linear likelihood transform (MLLT) [12] to conduct speaker-independent feature and model adaptation to train better acoustic models. Feature space maximum likelihood linear regression (FMLLR), also known as constrained maximum likelihood linear regression (CMLLR) [21], is a commonly adopted supervised adaptation method, where the feature transforms can be estimated at the speaker or utterance level using labeled data. With limited amounts of transcribed child speech, [13] improved acoustic models by freezing lower layers of the pre-trained DNN while only the output layer is updated. However, these methods usually require at least some quantity of human transcribed child speech for supervised training or fine-tuning, while in reality such annotated resources are publicly unavailable to most academic researchers. In addition, leveraging transcribed data from new domains to sequentially retrain pre-trained DNN models usually suffers from catastrophic forgetting [22, 23]. Such challenges motivate us to investigate unsuper-vised adaptation approaches.

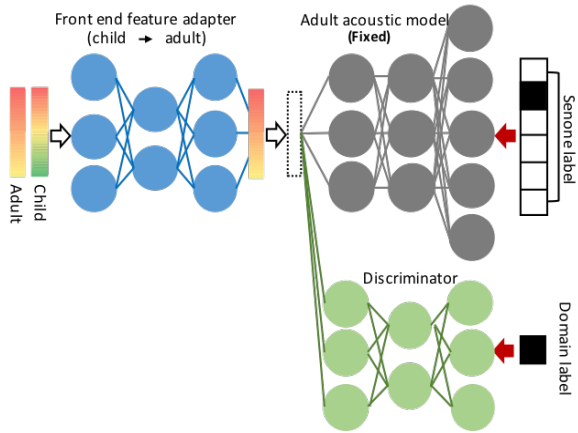

**Fig. 1**. Child-to-adult acoustic feature adaptation framework

## 2.2. Adversarial training

Inspired by generative adversarial networks (GAN), adversarial learning has been explored for learning domain-invariant models in areas such as image classification [15], robust speech recognition [16], speaker adaptation [17], and spoken keyword spotting [18]. Rather than training a domain-invariant acoustic model from scratch, [14] applied adversarial learning to explicitly estimate a global set of transformations to adapt child speech features to be used in a high performance adult acoustic model. Although it has achieved favorable performance, the implicit assumption that all phonemes/senones should be transformed in the same fashion could cast practical constraints in achieving robust performance. In this work, we strategically consider exploiting phonemic information to help anchor more targeted transformations at the senone level.

## 3. UNSUPERVISED FEATURE ADAPTATION VIA ADVERSARIAL MULTI-TASK LEARNING

Fig.1 shows the acoustic model architecture for processing child speech consists three sub-networks: an adult acoustic model, a child-to-adult adaptation model and a domain discriminator. Prior to model training, the parameters of the adult acoustic model are copied from a pre-trained adult model, which has been trained with large amounts of transcribed adult speech. These copied parameters are fixed during model training. We attach the output layer of the feature adaptation model to the input layer of the adult acoustic model. The discriminator is connected to the output layer of the front-end adaptation network, which is used to map the transformed features to domain labels. During inference, the domain discriminator component is removed; we only use adult acoustic model and the child to adult adaptation model to process child speech.

## 3.1. Feature adaptation with binary domain adversaries

The idea of domain adversarial learning is minimizing the domain discriminator prediction accuracy while training the generator to confuse the discriminator . In our previous work [14], during model training, the front-end feature adaptation network maximizes the loss of the child-adult binary domain classifier so the transformed features are mapped to become closer to the adult speech feature space. To ensure the generated features are able to perform senone classification, we minimized errors from acoustic modeling at the same time via multi-task training, where additional senone classification cross entropy loss is considered.

However, such binary domain adversaries suffer from a limitation: the feature adaptor aligns the two domains of child and adult from a global perspective. A global transformation is sub-optimal because the differences in acoustic variability of child and adult speech are different across phonemes. For example, there is less acoustic differences of child and adult speech for voiceless fricatives such as /f/ compared to the post-alveolar approximant /r/, as the cavity formed from the front of the lower lip to the upper teeth constriction when producing /f/ is similar in adults and children leading to similar resonant frequencies, while the the resonant frequencies caused by /r/ would differ in adults and children due to the size of the vibrating vocal folds, tongue position and oral cavity size [19]. Therefore, applying the same transformation to /f/ and /r/ is not ideal as /f/ requires minimal or no transformation while /r/ would require significant adaptation. Coarticulation effects could further complicate the acoustic variability across adults and children as well. Therefore, it is strategic to consider phonetic information such as senones (clustered triphones) when learning the adaptation transformations. Inspired by such linguistic insights, we exploit multi-task adversarial training to learn more robust transformations specific to each senone. Fig. 2 using an example of classifying two senones to illustrate the intuition behind the motivation of our current work. In Fig. 2(a), while after domain adaptation for child speech the two domains are closer to each other, the separation across senones is not demarcated, resulting in many senone samples falsely classified. Such limitations motivates us to incorporate senone knowledge into adversarial training to enable seonone-associated feature space transforms, which we elaborate in Section 3.2.

## 3.2. Senone-aware feature adaptation

As shown in Fig. 2(b), the binary child/adult domain information is now re-represented at the senone level. Therefore, the proposed representation not only embodies the broad domain information (adult and child speaker), but also encodes fine-grained senone information, enabling the proposed adaptation model to learn feature transforms specific to senones resulting in more distinct separation among senone classes.

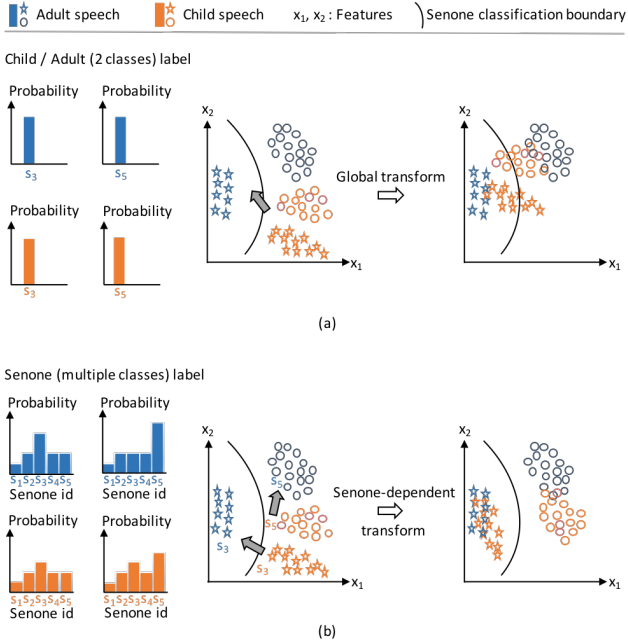

**Fig. 2**. (a) Global transformation using binary adversaril training (BAT): Only domain labels of *child* or *adult* are considered as the model is blind to senone information (e.g. $s_3$ and $s_5$ for adult speech are both labeled as adult domain). (b) Proposed senone-aware adversarial training (SAT): In addition to domain labels of child and adult, senone posteriors are also considered.

Assuming that there are $N$ training samples in total and $n$ samples of them belong to adult speech, the multi-task objective function is computed as:

$$E(\Theta_{adpt}, \Theta_{dom}) = \frac{1}{n}\sum_{i=1}^{n} L_{senone}^{i}(\Theta_{adpt}) - \frac{1}{N}\sum_{i=1}^{N} L_{dom}^{i}(\Theta_{adpt}, \Theta_{dom}) \quad (1)$$

where $L_{senone}$ is the cross-entropy loss between the adult senone labels and the network output posteriors from the adult acoustic model. $L_{dom}$ is the domain classification loss on adult and child speech samples.

The domain classification loss $L_{dom}$ in adversarial training using binary domain labels is characterized as:

$$L_{dom}^{i}(\Theta_{adpt}, \Theta_{dom}) = -(1 - I_{dom})\log P(dom = a \mid f_i) - I_{dom} \log P(dom = c \mid f_i) \quad (2)$$

where $I_{dom}$ is the domain indicator function, possessing the value of 1 for children samples $f_i$ and 0 for adult samples. $P(dom = a \mid f_i)$ and $P(dom = c \mid f_i)$ are the probability outputs from the domain discriminator.

To incorporate senone information, we present the domain loss $L_{dom}$ in Equation (2) as follows:

$$L_{dom}^{i} = -(1 - I_{dom})\sum_{k=1}^{K} \alpha_k^i \log P(dom = a, sen = k \mid f_i) \\ - I_{dom} \sum_{k=1}^{K} \alpha_k^i \log P(dom = c, sen = k \mid f_i) \quad (3)$$

where $K$ is the total number of senones. $\alpha_k^i$ is the $k$th entry of senone posteriors, extracted from the adult acoustic model for speech sample $i$. The child/adult domain information is thus represented by a $K$ dimension vector.

During model training, the parameters in the front-end feature adapter ($\Theta_{adpt}$) and domain classifier ($\Theta_{dom}$) are optimized such that:

$$\Theta_{adpt} = \underset{\Theta_{adpt}}{\operatorname{argmin}}\ E(\Theta_{adpt}, \Theta_{dom}) \quad (4)$$

$$\Theta_{dom} = \underset{\Theta_{dom}}{\operatorname{argmax}}\ E(\Theta_{adpt}, \Theta_{dom}) \quad (5)$$

Via joint optimization, the generated child speech features are not only closer to the adult speech features but are also able to contribute to sharper senone classification performance.

## 4. EXPERIMENTAL SETUP

### 4.1. Datasets

We adopt LibriSpeech [24] "train-clean-100" subset (251 speakers) to train the adult acoustic model. The *SingaKids-English* corpus [14] consists of 46 hrs from 193 speakers ranging from ages 6 to 12. The age and gender range for the developmental and test sets are equally distributed to minimize biased performance at test time. There are 5 proficiency levels used in assessment, which are scored by an English school teacher certified by the Ministry of Education in Singapore. We merge the six grades in primary school into three levels of G12, G34, and G56.

### 4.2. Implementation details

The training data for the adult acoustic model was parameterized into 40-dimensional log Mel-scale filter-bank features, along with their first and second temporal differentials. The input to the network are 11 contiguous frames that yield a 1,320-dimensional feature vector. The back-end adult acoustic model has 6 hidden layers with 2,048 nodes per layer. Both batch normalization and dropout were used to prevent over-fitting. Considering a language learner's pronunciation and fluency performance are often correlated, we designed a multi-task neural network architecture to perform the pronunciation and fluency assessment together. The input to the network is a 30-dimensional feature vector, consisting a set of features adopted from previous work [3, 25]. The model consists of 3 layers with 128 nodes per layer, and 2 softmax output layers with 5 nodes in each branch. All hyper-parameters were optimized on the developmental set of "Dev-clean" of LibriSpeech and the developmental set SingaKids-English.

Table 1. Phone error rate (%) on test set SingKids-English.

|  | Baseline | Baseline + adaptation | | |
|---|---|---|---|---|
|  | DNN | FMLLR | BAT[14] | SAT |
| G12 | 85.11 | 83.73 | 75.89 | 69.56 |
| G34 | 73.79 | 73.24 | 67.26 | 61.84 |
| G56 | 69.08 | 65.78 | 62.31 | 58.02 |
| Overall | 74.43 | 72.55 | 67.19 | 62.02 |

Table 2. Pronunciation prediction performance comparison.

|  | Baseline | BAT[14] | SAT |
|---|---|---|---|
|  | Accuracy(%) | | |
| G12 | 42.1 | 47.3 | 50.0 |
| G34 | 37.3 | 38.9 | 47.7 |
| G56 | 47.3 | 52.3 | 52.7 |
| Overall | 43.3 | 47.5 | 49.8 |
|  | MSE | | |
| G12 | 1.30 | 1.10 | 1.12 |
| G34 | 1.14 | 1.14 | 1.11 |
| G56 | 1.44 | 1.32 | 1.19 |
| Overall | 1.32 | 1.25 | 1.17 |

Table 3. Fluency prediction performance comparison.

|  | Baseline | BAT[14] | SAT |
|---|---|---|---|
|  | Accuracy(%) | | |
| G12 | 31.6 | 42.1 | 47.3 |
| G34 | 40.3 | 38.8 | 44.7 |
| G56 | 50.9 | 53.6 | 55.3 |
| Overall | 44.2 | 47.0 | 49.7 |
|  | MSE | | |
| G12 | 2.25 | 1.99 | 1.50 |
| G34 | 1.96 | 1.80 | 1.46 |
| G56 | 2.22 | 1.96 | 1.40 |
| Overall | 2.13 | 1.90 | 1.42 |

## 5. EXPERIMENTAL RESULTS

### 5.1. Speech recognition

To validate our DNN baseline that trained on adult speech, we employ the standard pruned version of the WSJ-5k trigram language model to conduct speech recognition on LibriSpeech "test-clean" set. Our baseline model achieves a word error rate of 9.49%, slightly better than Kaldi's DNN model[1] of 9.66%, suggesting this baseline is competitive. Other baselines include a lightly supervised FMLLR and the domain adaptation using binary adversarial training (BAT) [14]. The FMLLR transformation matrix was estimated with phonetic labels generated from the aforementioned DNN baseline model. (Note that we focus on investigating unsupervised approaches, dedicated methods for post-processing those generated labels for FMLLR are out of the scope of this paper.) To minimize the influence from the language model, we adopt a free phone decoding graph instead.

As shown in Table 1 the proposed senone-aware approach (SAT) achieves the lowest phone error rate (PER) in all conditions. It reduces the overall PER for 12.41% absolute over the DNN baseline, and 7.7% relative over the BAT baseline. Though the PER around 60% is still relatively high, such results are consistent with prior work [13, 26], and illustrate the challenges of processing child speech.

### 5.2. Pronunciation and fluency evaluation

We use two widely adopted metrics [8, 14]: prediction accuracy and mean squared error (MSE) between the predicted scores and the human rated scores. From Table 2 and 3, we obverse that the proposed SAT approach improves the prediction accuracy on both pronunciation and fluency evaluation, implying SAT generates features that are more easily separable for the scoring classifier. The prediction accuracy is around 50%, more than twice the probability at chance level. The overall MSE is reduced from 1.90 to 1.42, 25.2% relative improvement when predicting fluency scores. Note that fluency scoring is done at the utterance level, so the available training data for the assessment classifier is much fewer than that for acoustic modeling, limiting the prediction performance. Approaches to relieve the need of such labor-intensive human scoring is a topic of on-going research.

[1]https://github.com/kaldi-asr/kaldi/blob/master/egs/librispeech/s5

## 6. CONCLUSION

To tackle challenges arising from the limited linguistic resources suitable for modeling child speech, we proposed an unsupervised adversarial multi-task training approach to transform child speech features to the adult feature space by anchoring the adaption at the senone-level to exploit senone-specific acoustic variations across adults and children. We showed that our approach outperforms competitive SOTA baselines by 7.7%-25.2% relative across three speech tasks.

# 7. REFERENCES

[1] S. Ghai and R. Sinha, "Exploring the role of spectral smoothing in context of children's speech recognition," in *Tenth Annual Conference of the International Speech Communication Association*, 2009.

[2] H. Liao, G. Pundak, O. Siohan, M. K. Carroll, N. Coccaro, Q.-M. Jiang, T. N. Sainath, and Senior, "Large vocabulary automatic speech recognition for children," in *INTERSPEECH*, 2015, pp. 1611–1615.

[3] C. Cucchiarini, H. Strik, and L. Boves, "Automatic evaluation of dutch pronunciation by using speech recognition technology," in *ASRU*. IEEE, 1997, pp. 622–629.

[4] H. Franco, V. Abrash, K. Precoda, H. Bratt, R. Rao, J. Butzberger, R. Rossier, and F. Cesari, "The SRI EduSpeakTM system: Recognition and pronunciation scoring for language learning," *Proceedings of InSTILL*, pp. 123–128, 2000.

[5] J. Cheng and J. Shen, "Towards accurate recognition for children's oral reading fluency," in *2010 IEEE Spoken Language Technology Workshop*, 2010, pp. 103–108.

[6] A. Metallinou and J. Cheng, "Using deep neural networks to improve proficiency assessment for children english language learners," in *INTERSPEECH*, 2014.

[7] Y. Qian, X. Wang, and K. Evanini, "Self-adaptive dnn for improving spoken language proficiency assessment.," in *INTERSPEECH*, 2016, pp. 3122–3126.

[8] K. Kyriakopoulos, M. Gales, and K. Knill, "Automatic characterisation of the pronunciation of non-native english speakers using phone distance features," 2018.

[9] A. Potamianos, S. Narayanan, and S. Lee, "Automatic speech recognition for children," in *EUSPEECH*, 1997.

[10] M. Gerosa, D. Giuliani, and F. Brugnara, "Acoustic variability and automatic recognition of children's speech," *Speech Communication*, vol. 49, pp. 847–860, 2007.

[11] M. Russell and S. D'Arcy, "Challenges for computer recognition of children's speech," in *Workshop on Speech and Language Technology in Education*, 2007.

[12] P. G. Shivakumar, A. Potamianos, S. Lee, and S. Narayanan, "Improving speech recognition for children using acoustic adaptation and pronunciation modeling.," in *WOCCI*, 2014, pp. 15–19.

[13] M. Matassoni, R. Gretter, D. Falavigna, and D. Giuliani, "Non-native children speech recognition through transfer learning," in *ICASSP*. IEEE, 2018, pp. 6229–6233.

[14] R. Duan and N. F. Chen, "Unsupervised feature adaptation using adversarial multi-task training for automatic evaluation of children's speech," in *INTERSPEECH*, 2020, pp. 3037–3041.

[15] Y. Ganin, E. Ustinova, H. Ajakan, P. Germain, H. Larochelle, F. Laviolette, M. Marchand, and V. Lempitsky, "Domain-adversarial training of neural networks," *JMLR*, vol. 17, no. 1, pp. 2096–2030, 2016.

[16] Y. Shinohara, "Adversarial multi-task learning of deep neural networks for robust speech recognition.," in *INTERSPEECH*, 2016, pp. 2369–2372.

[17] Z. Meng, J. Li, Z. Chen, Y. Zhao, V. Mazalov, Y. Gang, and B.-H. Juang, "Speaker-invariant training via adversarial learning," in *ICASSP*, 2018, pp. 5969–5973.

[18] J. Hou, P. Guo, S. Sun, F. K. Soong, W. Hu, and L. Xie, "Domain adversarial training for improving keyword spotting performance of esl speech," in *ICASSP*. IEEE, 2019, pp. 8122–8126.

[19] K. N. Stevens, *Acoustic phonetics*, vol. 30, MIT press, 2000.

[20] R. Serizel and D. Giuliani, "Vocal tract length normalisation approaches to dnn-based children's and adults' speech recognition," *IEEE SLT*, pp. 135–140, 2014.

[21] M. J. Gales and P. C. Woodland, "Mean and variance adaptation within the mllr framework," *Computer Speech & Language*, vol. 10, no. 4, pp. 249–264, 1996.

[22] R. M. French, "Catastrophic forgetting in connectionist networks," *Trends in cognitive sciences*, vol. 3, no. 4, pp. 128–135, 1999.

[23] S.-W. Lee, J.-H. Kim, J. Jun, J.-W. Ha, and B.-T. Zhang, "Overcoming catastrophic forgetting by incremental moment matching," in *Advances in Neural Information Processing Systems 30*, I. Guyon, U. V. Luxburg, S. Bengio, H. Wallach, R. Fergus, S. Vishwanathan, and R. Garnett, Eds., pp. 4652–4662. 2017.

[24] V. Panayotov, G. Chen, D. Povey, and S. Khudanpur, "Librispeech: an asr corpus based on public domain audio books," in *ICASSP*. IEEE, 2015, pp. 5206–5210.

[25] W. Hu, Y. Qian, F. K. Soong, and Y. Wang, "Improved mispronunciation detection with deep neural network trained acoustic models and transfer learning based logistic regression classifiers," *Speech Communication*, vol. 67, pp. 154–166, 2015.

[26] R. Serizel and D. Giuliani, "Deep-neural network approaches for speech recognition with heterogeneous groups of speakers including children," *Natural Language Engineering*, vol. 23, no. 3, pp. 325–350, 2017.